\documentclass[onecolumn,prd,superscriptaddress,showpacs,floatfix,%
preprintnumbers,nofootinbib,showkeys]{revtex4}  

\usepackage{amsmath,amsfonts}
\usepackage{graphicx}
\usepackage{epsfig,epsf}

\def\bar {\overline}

\def\be {\begin{equation}}
\def\ee {\end{equation}}
\def\beq {\begin{equation}}
\def\eeq {\end{equation}}
\def\bea {\begin{eqnarray}}
\def\eea {\end{eqnarray}}

\def\bra {\langle}
\def\ket {\rangle}

\def\beq{\begin{equation}}
\def\eeq{\end{equation}}
\def\barr{\begin{array}}
\def\earr{\end{array}}

\def\opcit(#1){ {\em op. cit.}, #1}

\def\issue(#1,#2,#3){#1, #2 (#3)} 

\def\APP(#1,#2,#3){Acta Phys.\ Polon.\ \issue(#1,#2,#3)}
\def\ARNPS(#1,#2,#3){Ann.\ Rev.\ Nucl.\ Part.\ Sci.\ \issue(#1,#2,#3)}
\def\CPC(#1,#2,#3){Comp.\ Phys.\ Comm.\ \issue(#1,#2,#3)}
\def\CIP(#1,#2,#3){Comput.\ Phys.\ \issue(#1,#2,#3)}
\def\EPJC(#1,#2,#3){Eur.\ Phys.\ J.\ C\ \issue(#1,#2,#3)}
\def\EPJD(#1,#2,#3){Eur.\ Phys.\ J. Direct\ C\ \issue(#1,#2,#3)}
\def\IEEETNS(#1,#2,#3){IEEE Trans.\ Nucl.\ Sci.\ \issue(#1,#2,#3)}
\def\IJMP(#1,#2,#3){Int.\ J.\ Mod.\ Phys. \issue(#1,#2,#3)}
\def\JHEP(#1,#2,#3){J.\ High Energy Physics \issue(#1,#2,#3)}
\def\JPG(#1,#2,#3){J.\ Phys.\ G \issue(#1,#2,#3)}
\def\MPL(#1,#2,#3){Mod.\ Phys.\ Lett.\ \issue(#1,#2,#3)}
\def\NP(#1,#2,#3){Nucl.\ Phys.\ \issue(#1,#2,#3)}
\def\NIM(#1,#2,#3){Nucl.\ Instrum.\ Meth.\ \issue(#1,#2,#3)}
\def\PL(#1,#2,#3){Phys.\ Lett.\ \issue(#1,#2,#3)}
\def\PRD(#1,#2,#3){Phys.\ Rev.\ D \issue(#1,#2,#3)}
\def\PRL(#1,#2,#3){Phys.\ Rev.\ Lett.\ \issue(#1,#2,#3)}
\def\SJNP(#1,#2,#3){Sov.\ J. Nucl.\ Phys.\ \issue(#1,#2,#3)}
\def\ZPC(#1,#2,#3){Zeit.\ Phys.\ C \issue(#1,#2,#3)}

\begin{document}

\title{
Controlling the fine-tuning problem with singlet scalar dark matter
}

\author{Indrani Chakraborty}

\author{Anirban Kundu}
\affiliation{Department of Physics, University of Calcutta,\\
92, Acharya Prafulla Chandra Road, Kolkata 700009, India. }

\date{\today}

\begin{abstract} 

Assuming that no other conventional new physics is found immediately at the LHC, we investigate how 
just the consistent solution of the scalar mass hierarchy problem points towards the minimal necessary field content.
We show that to ameliorate the fine-tuning problem, one needs to introduce more scalar degrees of 
freedom. The simplest solution is one or more real singlets (with the possibility of combining two of them 
in a complex singlet), which may act as viable cold dark matter candidates, because the constraints on 
the scalar potential disfavor 
any mixing between the new scalar(s) with the SM doublet. Furthermore, the fine-tuning problem of 
the new scalars necessitates the introduction of vector-like fermions. Thus, singlet scalar(s) and vector 
fermions are minimal enhancements over the Standard Model to alleviate the fine-tuning problem.
We also show that the model predicts Landau poles for all the scalar couplings, whose positions depend 
only on the number of such singlets. Thus, introduction of some new physics at that scale becomes inevitable.
We also discuss how the model confronts the LHC constraints and the latest XENON100 data.  

\end{abstract}



\maketitle

\section{Introduction}

Assuming that the Higgs boson has been found at $m_h = 125.9\pm 0.4$ GeV 
\cite{cms-higgs,atlas-higgs}, the Standard Model (SM) is now complete, 
but there are enough reasons to believe that this is not the final story, rather,
at best an effective theory valid up to a certain energy scale. Two
such major reasons are the fine-tuning problem of the Higgs boson mass and the existence 
of the dark matter (DM). One might even add the massive neutrinos.

There are various extensions of the SM to explain the first problem; some of them,
like R-parity conserving supersymmetry, Universal extra dimensions, or Little Higgs models 
with T-parity (i.e. almost any model with a discrete $Z_2$ symmetry), 
also provide us a viable DM candidate.
Unfortunately, till now there has been no direct detection in the colliders 
of any new physics (NP) particle. While there are some interesting hints, there are
no compelling indirect evidence too, coming from low-energy observables like those
in flavour physics. 

Recently, it was pointed out in Refs.\ \cite{degrassi,alekhin} that the Higgs boson quartic 
coupling, $\lambda$, has such a value for $m_h \approx 125$ GeV that it might remain 
perturbative all the way up to the Planck scale; neither does it blow up and hit the 
Landau pole, nor does it become negative and make the electroweak vacuum unstable. Thus, 
there seems to be a distinct possibility that a desert lies between the electroweak scale,
parametrized by $v$ $(\approx 246$ GeV), the vacuum expectation value (VEV) of the Higgs field,
and the Planck scale $M_{Pl}$ $(\sim 10^{19}$ GeV). 

In this paper, we would like to investigate whether such a possibility is allowed without 
running afoul of the fine-tuning problem, and if not, what can the possible directions 
of solution be. It was pointed out long ago by Veltman \cite{veltman} that the fine-tuning 
problem can be ameliorated if the quadratically divergent term for Higgs self-energy, generated 
by all the degrees of freedom of the model, is either zero or very small, by some 
symmetry of the model. For example, all supersymmetric models are only a particular 
application of this idea, where the contributions from bosonic and fermionic degrees of 
freedom cancel in the limit of exact supersymmetry. 

It is a well-known fact that the SM falls short of the Veltman condition (VC), now that 
we know all the masses. Thus, there must be some new degrees of freedom lurking somewhere, which
interact with the Higgs boson and contribute to its self-energy. If we apply the principle of
Occam's razor and focus only on the fine-tuning problem, we will see that the simplest 
such extension is to introduce one or more scalars; more chiral fermions only worsen the fine-tuning.
We will concentrate on gauge singlet scalars, whose phenomenology is well-investigated 
\cite{barger-real,barger-cmplx,batell1,belanger,other-singlet}, including the LHC constraints 
\cite{mambrini}, and the stability of the electroweak vacuum \cite{kadastik}. The role of singlet scalars 
for solving the fine-tuning problem was discussed earlier in Refs.\ \cite{aksrc,grzad1} and has recently
been investigated in Ref.\ \cite{drozd}. It was shown in Ref.\ \cite{moresinglets} that with a 
large number of singlets, the electroweak phase transition may become a strong first-order one.
As is well-known, singlet scalars with a right-sign mass term, and therefore without a VEV,
do not mix with the SM doublet, and thus can be good candidates for cold DM (see Section II for details)
provided they satisfy other constraints like the DM density or DM-nucleon scattering cross-section limits.
However, the fine-tuning problem is not generally addressed in the scalar DM models, assuming 
that some other mechanism must take care of the same. An exception is Ref.\ \cite{bazzocchi}, 
where the interplay of the Veltman condition (for the SM Higgs only) and scalar DM was discussed.
We show how the fine-tuning problem can be interlinked with one or more possible DM candidates, 
treating all the Veltman conditions consistently. 

We will show in this paper that the solution of the VC with one or more singlets disfavors 
the scenario where the singlets mix with the SM doublet. In that case, they can be good cold DM 
candidates. At the same time, the solution of the VC for the singlets necessitates the introduction
of vectorial fermions. However, they are much less constrained from precision electroweak 
observables than new chiral fermions (degenerate vector fermions that do not mix with the SM 
chiral fermions do not contribute to the oblique parameters $S$ and $T$) \cite{cynolter}, 
and from the rate of Higgs production via gluon-gluon
fusion as they do not couple with the SM Higgs. Also, a new heavy chiral generation would make the 
Higgs self-coupling $\lambda$ to go negative and make the vacuum unstable, but vector fermions,
which are either only singlets or only doublets,
do not run into that problem as they couple with the doublet Higgs if and only if they 
mix with the SM fermions. The role of vector fermions, when they mix with the SM fermions, or when the 
125 GeV scalar is an admixture of a doublet and a singlet state, in enhancing the $h\to\gamma\gamma$ rate 
has been investigated in the literature \cite{vf-2gamma}. Talking about the diphoton excess, 
it has been shown that if this anomaly 
persists, it will point to some new bosonic degrees of freedom at a few TeV
\cite{nima}. Furthermore, these fermions can play an important 
role to make the electroweak phase transition a strong first-order one \cite{carena,davoudiasl}. 
However, note that neither the diphoton decay rate nor the evolution of the Higgs quartic coupling 
is affected if the new vectorial fermions fall in a definite
representation of SU(2) (only singlets or only doublets), and there is no mixing with the SM fermions.
Spin-$\frac12$ DM candidates and possible allowed effective operators have also received attention
\cite{cheung}, but here we do not consider the possibility that the vector fermions are the DM candidates.

We will next concentrate on the evolution of the new couplings with energy. We will show that while 
the Yukawa couplings of the new fermions remain under control, the scalar self couplings tend
to blow up and hit the Landau pole. The exact position of the 
pole depends on the number of singlets $N$, which is less than 1 TeV for one singlet, about 
150 TeV for $N=4$, and more than $10^4$ TeV for $N=10$. Of course, some way before the Landau pole, the 
couplings cease to be 
perturbative. If we take this scale as the point of onset of NP, we have every reason to hope
that some new degrees of freedom will show up at the upgraded LHC at $\sqrt{s}=14$ TeV unless 
the number of such singlets is not abnormally large. This need not be in the form of the 
gauge singlet scalars, which will be hard to produce at the LHC, but possibly be in the form 
of new vector fermions, where the colored fermions can be produced from gluon-gluon or quark-antiquark 
fusion. However, this statement requires a more careful discussion, to which we will come back
later.

Thus, we can rephrase our aim: if one does not see supersymmetry, or extra dimensional models, or 
any other popular NP scenarios at the LHC, does it 
necessarily mean the possibility of 
a barren desert between the electroweak scale and the Planck scale? The answer, as expected, is
in the negative, but a spin-off is a possible candidate for cold DM.  
Solution of the fine-tuning problem hints 
at bosonic DM, for which the singlet scalars are perhaps the best candidate. We show that a theory with 
only SM augmented by singlet scalar(s) necessarily mean some NP at the TeV scale or higher. 
To put it in another way, if one considers the singlet scalar DM scenario with an eye to the 
fine-tuning problem of the scalar masses, there exists a solution but only with the introduction 
of vectorial fermions; this is the minimal enhancement over the SM that is necessary.

The paper is arranged as follows. In Section II, we discuss the scalar potential and the corresponding 
Veltman conditions. In Section III, we study the renormalization group (RG) evolution of the couplings
and its possible consequences. We summarize and conclude in Section IV.

\section{The Veltman Condition}
  
Let us first consider the SM augmented with one real singlet scalar $S$.
The potential is
\be
V(\Phi,S) = V_{\rm SM} + V_{\rm singlet}
= -\mu^2\Phi^\dag\Phi + \lambda (\Phi^\dag\Phi)^2 
-M^2 S^2 + \tilde\lambda S^4 + a S^2(\Phi^\dag\Phi)\,,
\label{onesing}
\ee
where $\Phi$ is the SM doublet, with $\bra\Phi\ket = v/\sqrt{2}$. We will take both 
$\mu^2, M^2 > 0$ to start with, and denote the
SM Higgs boson, the remnant of $\Phi$ after spontaneous symmetry breaking, by $h$. 
There might be a cubic term $cS^3$ in the potential, but that would not affect the 
subsequent analysis. The linear terms in $S$ giving rise to tadpole diagrams are 
assumed to cancel out and they will remain so even after the quantum corrections.  
This happens if we take the tadpole potential to be \cite{barger-real} 
\be
V_{\rm tadpole} = \alpha_1 S + \alpha_2 \Phi^\dag\Phi S
\ee
with $\alpha_1 +\frac12\alpha_2 v^2 = 0$. 
There might also be a discrete symmetry, like $S\to -S$, preventing odd terms. 
For $N$ singlets with an $O(N)$ symmetry, the potential looks like 
 \be
V(\Phi,S_i) = -\mu^2\Phi^\dag\Phi + \lambda (\Phi^\dag\Phi)^2 
-M^2 \sum_i S_i^2 + \tilde\lambda \left(\sum_i S_i^2\right)^2 + a (\Phi^\dag\Phi)\sum_i S_i^2\,.
\label{multising}
\ee

With only the SM, i.e., $a=0$, the Higgs self-energy receives a quadratically 
divergent correction
\be
\delta m_h^2 = \frac{\Lambda^2}{16\pi^2} \left(6\lambda + \frac34 g_1^2 + \frac94 g_2^2
- 6 g_t^2\right)\,,
\label{smvc}
\ee
where $g_1$ and $g_2$ are the $U(1)_Y$ and $SU(2)_L$ gauge couplings, and $g_t = \sqrt{2}
m_t/v$ is the top quark Yukawa coupling. We treat all other fermions as massless, and use 
the cut-off regularization, $\Lambda$ being the cutoff scale. Note that while this 
regularization is not Lorentz invariant, this has the nice feature of separating the 
quadratic and the logarithmic divergences. Other methods, such as dimensional 
regularization, do not discriminate between these two divergences, and we get a
slightly different correction \cite{einhorn}, which includes both quadratic and logarithmic 
divergences:
 \be
\delta m_h^2 \propto \frac{1}{\epsilon} \left(6\lambda + \frac14 g_1^2 + \frac34 g_2^2
- 6 g_t^2\right)\,.
\ee
As our goal is to cancel the strongest divergence, we will use the cut-off regularization.

The Veltman condition (VC) demands that the quantity inside parentheses in Eq.\ (\ref{smvc})
be made zero, or at least controllably small,
 by some symmetry. There are further quadratic divergences coming from two-loop 
diagrams, but they are suppressed from one-loop contributions by a factor of 
$\ln(\Lambda/\mu)/16\pi^2$, where $\mu$ is the regularization scale, and is in general under 
control. 

One can say that the quadratic divergence is under control if $|\delta m_h^2| \leq m_h^2$, 
which translates into
\be
\left\vert m_h^2 + 2m_W^2 + m_Z^2 - 4m_t^2\right\vert \leq \frac{16\pi^2}{3} \frac{v^2}{\Lambda^2} m_h^2\,.
\label{smvc2}
\ee
This inequality is clearly not satisfied in the SM for $v^2/\Lambda^2 \leq 0.1$, or $\Lambda \geq 760$ GeV,
and onset of NP at such a low scale is almost ruled out by the LHC. 
Any more chiral fermions will only 
make the left-hand side of Eq.\ (\ref{smvc2}) more negative and hence worsen the fine-tuning
problem. What is needed is a positive contribution, and extra scalars are a viable option. 

With one extra singlet $S$, as in Eq.\ (\ref{onesing}), the VC is modified to
\be
\delta m_h^2 = \frac{\Lambda^2}{16\pi^2} \left(6\lambda + \frac34 g_1^2 + \frac94 g_2^2
- 6 g_t^2 + a\right)\,,
\label{singletvc1}
\ee
and with $N$ number of identical singlets (i.e., an $O(N)$ symmetric singlet sector), 
the last term is replaced by $Na$. 

For $N=1$, we find that $a = 4.17$, which is quite large even if not nonperturbative 
(we take $4\pi \approx 12.56$ to be the threshold for nonperturbativity to set in). More singlets 
bring down the value to $4.17/N$. This is to be taken as an indicative value only, as there
is no reason why this value would be absolutely stable if we take higher-order corrections
(for an estimate of higher-order effects, we refer the reader to \cite{drozd}, where it can be seen 
that such corrections bring a marginal change.) To be consistent, we will limit our discussions 
within one-loop effects only, and therefore use only one-loop renormalization group (RG) 
equations to calculate the evolution of the couplings.

We need to consider the VC for the singlet too. The condition does not depend on whether 
the singlet develops a VEV or not. This reads
\be
\delta m_S^2 = \frac{\Lambda^2}{16\pi^2}[(8+4N)\tilde\lambda + 4a]\,.
\label{singletvc2}
\ee
So, only with singlet, we need a large (and definitely nonperturbative) and negative quartic 
coupling, and the potential develops a minimum unbounded from below in the direction 
$|\Phi| = {\rm constant}$ and $|S|\to \infty$. Thus, this solution is clearly unacceptable. 

While one needs some negative contribution to Eq.\ (\ref{singletvc2}), one notes that this cannot 
come from chiral fermions of the SM as they do not couple to $S$. Thus, one is led to introduce 
vector fermions --- either singlets or doublets under $SU(2)$. This introduces further terms in the 
potential:
\be
{\cal L}_{VF} = -m_F \bar{F}F - \zeta_F \bar{F} F S\,,
\ee
and the mass of $F$ is $m_F + \zeta_F\bra S \ket$. Note that a symmetry like $S\to -S$ implies
$F\to i\gamma_5F$ and hence forbids the bare mass term, unless the symmetry is explicitly broken. 
We will not pursue this possibility further as we show explicitly that a nonzero VEV to the 
singlet is disfavored, so the vector fermions must get their masses from the bare mass term.
For this case, {\em i.e.},  $M^2 < 0$ and 
$\bra S \ket = 0$, there is no mass term from the Yukawa couplings. Direct searches at the LHC 
put a lower limit of the order of 500 GeV on the mass of vector quarks.  

For simplicity, we assume a complete generation of vector fermions $(N,E),~(U,D)$ with all Yukawa
couplings $\zeta_i$ to be the same (the heavy neutrino $N$ is not to be confused with $N$, the number of
singlets). One can, in principle, consider only one such fermion in the 
spectrum, or only the lepton or quark doublet. The VC for $S$ now reads
 \be
\delta m_S^2 = \frac{\Lambda^2}{16\pi^2}[(8+4N)\tilde\lambda + 4a - 4Z^2]\,,
\label{singletvc3}
\ee   
where
\be
Z^2 = \sum_i N_c  \zeta_i^2 = \zeta_E^2 + \zeta_N^2 + 3(\zeta_U^2+\zeta_D^2) = 8\zeta^2\,,
\ee 
$N_c$ being the color of the corresponding fermions. 
While this does not guarantee a degenerate generation, that remains a distinct possibility, 
and thus one can avoid the strong constraints coming from oblique $S$ and $T$ parameters---because
of the vectorial nature and degeneracy.  
  
If there is only one singlet and $M^2 > 0$, the minimization conditions are 
\be
-\mu^2 + \lambda v^2 + a {v'}^2 = 0\,,\ \ \ 
-M^2 + 2\tilde\lambda {v'}^2 + \frac12 av^2 = 0\,,
\ee
where $\bra S \ket = v'$. The mass term can be written as
\be
\begin{pmatrix}h & S\end{pmatrix}
 {\cal M} \begin{pmatrix} h\cr S\end{pmatrix}
= \begin{pmatrix} h & S\end{pmatrix} 
\begin{pmatrix}\lambda v^2 & avv'\cr
avv' & 4\tilde\lambda {v'}^2\end{pmatrix}
 \begin{pmatrix} h \cr S\end{pmatrix}\,.
\ee
The condition for both masses to be real is 
\be
4 \lambda \tilde\lambda \geq a^2\,.
\ee
For $\lambda \approx 0.13$ and $a \approx 4$, this makes $\tilde\lambda$ very large ($\geq 33$) 
and clearly nonperturbative. While this by itself may still be acceptable, the fact that 
all the scalar couplings hit their respective Landau poles almost right at the electroweak scale rules 
this option out.

One might wonder whether the situation improves in the large-$N$ limit. However, if the original scalar
sector has an $O(N)$ symmetry, a spontaneous symmetry breaking will result in $N-1$ Goldstone bosons,
which couple to the doublet Higgs $h$, and therefore will give a very large invisible decay width of $h$. 
This is again unacceptable from the measurement of the branching fractions of the Higgs at the LHC.  

Thus, we are forced to take $M^2 < 0$, so that the singlet does not develop a VEV and there is no 
singlet-doublet mixing. This is true even if there are more than one such singlets. 
The mass-squared of the singlet is given by
\be
m_S^2 = 2M^2 + av^2\,,
\ee
so for $N=1$, the lowest possible mass for the singlet is about 500 GeV, and goes down as $1/\sqrt{N}$. 
This means no change in the decay pattern of the 125 GeV scalar from the 
SM Higgs, and no $h\to SS$ invisible decay unless $N$ is so large that $m_S < m_h/2$ (this happens 
for $N\geq 66$). Also, there are no changes to the 
oblique parameters $S$, $T$, and $U$ \cite{barger-real}. 

\subsection{The Cosmological Constraints}

Although the mass of the singlet(s) has a lower limit, it is nevertheless a free parameter 
of the theory. The number of singlets $N$ is another free parameter, but the singlet-doublet 
coupling strength $a$ is not (assuming an $O(N)$ symmetric singlet sector). The $O(N)$ symmetry 
might be broken softly and by a very small amount to make the lightest scalar nondegenerate. 
The vector fermions have only a marginal influence on the viability of this singlet as a 
DM candidate if $m_F > m_S$, which we will assume to be true in our entire analysis.

\begin{figure}
\vspace{-15pt}
\centerline{
\rotatebox{270}{\epsfxsize=5cm\epsfbox{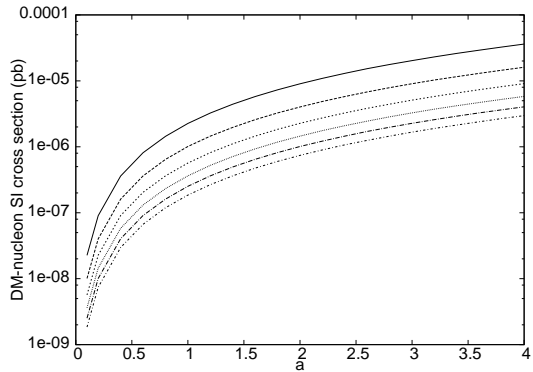}}
}
\caption{The spin-independent dark matter-nucleon cross section as a function of the singlet-doublet 
coupling parameter $a$. The lines, from top to bottom, are for the singlet mass $m_S = 200,
300,400,500,600$ and 700 GeV respectively. Even for $N$ such singlets, we have assumed a slight 
nondegeneracy (see text) so that only one of the singlets is the DM candidate.}
   \label{fig1}
\end{figure}

Over almost the entire parameter space relevant for our discussion (100 GeV $< m_S<$ 800 GeV,
$0.1 < a < 4.2$), the singlet density is below the DM relic density \cite{wmap9}
\be
\Omega_{DM}h^2 = 0.1138\pm 0.0045\,,
\ee
except for high $m_S$ and very low $a$, so there is no conflict with overclosure of the universe.
The singlet can scatter off the nucleons, by exchange of the Higgs boson; it can be a tree-level 
scattering off the quarks or a one-loop scattering off the gluons. The cross-section is necessarily 
spin-independent, which is rather strongly bounded by the  
data from 225 days running of XENON100 \cite{xenon100}. Roughly, the $2\sigma$ upper bound for
spin-independent DM-nucleon scattering is of the order of 2-4$\times 10^{-8}$ pb, for the DM mass
between 300 and 1000 GeV. We have checked through the software micrOMEGAs \cite{micromegas} that 
this bound is satisfied only for small couplings ($a \leq 0.3$-0.5, depending on $m_S$), see 
Fig.~\ref{fig1} for details. This is 
easy to understand as the cross-section grows as $a^2$. This might mean that for a successful 
solution of the fine-tuning problem, we need a large number of scalars $N\sim 10$, leading to the 
possibility of a strong first-order electroweak phase transition. However, 
around a mass window of 65-100 GeV, where $h \to SS$ would still be forbidden, there is an allowed
window with a much larger DM-nucleon spin-independent cross-section found by Dama/I \cite{dama}
(this is in apparent conflict with XENON100), where one can have larger values of $a$ and hence 
a smaller number of singlets.  

\subsection{The vector fermions}

The vector fermions also merit a brief discussion. First, let us note that they have to be vectorial 
because (i) they must couple to the singlet, and (ii) heavy chiral fermions are highly constrained from
the LHC data, and they drive the SM Higgs quartic coupling $\lambda$ to negative values and hence 
make the electroweak vacuum unstable. The vector fermions can be $SU(2)$ singlets or doublets; the only difference 
is their couplings to the electroweak gauge bosons. 

Second, We have assumed a complete generation of $(N,E)$
and $(U,D)$, but one can work with partial generations too. In fact, just a single vector neutrino 
would have been sufficient to solve the Veltman condition for the singlet scalar, albeit with a 
much higher Yukawa coupling. If all the vector fermions are exactly degenerate (which is
not necessary for the model, but just a simplifying assumption) stationary states,
they would have been stable, which is in conflict with cosmological bounds on stable charged and/or 
colored particles. A possible way out would be a very small admixture of the vector fermions, at least
the lightest lepton and quark if we admit a nondegeneracy, with the chiral fermions. With such a tiny 
admixture, the vector fermions will be stable on the collider scale but not on the cosmological scale.
At the same time, none of the flavor observables would be significantly affected.  
The scalar $S$ can be a DM candidate if $m_S < m_F$, which we assume in our analysis.

The vector quarks can be produced at the LHC by $gg$ or $q\bar{q}$ fusion, just like a heavy sequential
quark. They will leave a jetty track by radiating off gluons, and will ultimately dump their 
energy in the hadron calorimeter. In fact, if they are stable on the collider scale (which is a 
necessity if the singlet scalar has to be a DM candidate --- otherwise the scalar will decay to 
lighter SM states through virtual vector fermions), 
they should form bound states with ordinary quarks. 
LHC experiments have searched for such long-lived quarks or leptons \cite{atlas1211}, and the mass limits 
depend on the quantum numbers of them. The lower limit for the mass of a long-lived lepton is of the order of 350-400 GeV, and that of a quark which can form a bound state is of the order of 700 GeV. If they are close to
the mass limits, LHC at $\sqrt{s}=14$ TeV might be able to produce them.
The vector fermions must also be almost degenerate so as not to affect the oblique parameters.   

One should also note that ultraheavy fermions run the risk of having their Yukawa couplings hitting the 
Landau pole \cite{chanowitz}. However, here the mass does not come from the Yukawa couplings but from the 
bare mass terms, so such a constraint is not applicable here.

\section{The one-loop RG equations}

Next, we would like to see how the couplings evolve with energy. The one-loop $\beta$-functions are 
\bea
16\pi^2 \beta_\lambda &=& 12\lambda^2 + 6g_t^2\lambda + N a^2 -\frac32\lambda(g_1^2+3g_2^2) 
- 3 g_t^4 + \frac{3}{16}(g_1^4 + 2 g_1^2 g_2^2 + 3 g_2^4)\,,\nonumber\\
16\pi^2 \beta_{\tilde\lambda} &=& (32 + 4N) {\tilde\lambda}^2 + a^2 + 4\tilde\lambda Z^2 - \sum_i N_c \zeta_i^4
\,,\nonumber\\
16\pi^2 \beta_a &=& \left[ 6\lambda + 12\tilde\lambda + 4a + 6g_t^2 + 4Z^2 -\frac32 g_1^2 -\frac92 g_2^2\right]a
\,,\nonumber\\
16\pi^2 \beta_{g_t} &=& \left[ \frac94 g_t^2 - \frac{17}{24}g_1^2 - \frac98 g_2^2 - 4g_3^2\right] g_t\,,
\nonumber\\
16\pi^2\beta_{g_3} &=& -\frac{17}{6} g_3^3 \theta(Q^2-m_U^2) - \frac{19}{6} g_3^3 \theta(m_U^2-Q^2)\,,\nonumber\\
16\pi^2 \beta_{\zeta_U} &=& \left[ \frac32 \zeta_U^2 + Z^2 -\frac43\left(\frac{1}{12}\right) g_1^2 - 0
\left(\frac{9}{4}\right) g_2^2 - 4g_3^2\right]\zeta_U\,,\nonumber\\
16\pi^2 \beta_{\zeta_D} &=& \left[ \frac32 \zeta_U^2 + Z^2 -\frac13\left(\frac{1}{12}\right) g_1^2 - 0
\left(\frac{9}{4}\right) g_2^2 - 4g_3^2\right]\zeta_D\,,\nonumber\\
16\pi^2 \beta_{\zeta_N} &=& \left[ \frac32 \zeta_N^2 + Z^2 -0\left(\frac{3}{4}\right) g_1^2 - 0
\left(\frac{9}{4}\right) g_2^2 \right]\zeta_N\,,\nonumber\\
16\pi^2 \beta_{\zeta_E} &=& \left[ \frac32 \zeta_E^2 + Z^2 -3\left(\frac{3}{4}\right) g_1^2 - 0
\left(\frac{9}{4}\right) g_2^2 \right]\zeta_E\,,
\label{all-rge}
\eea
where $\beta_h\equiv dh/dt$, and $t \equiv \ln(Q^2/\mu^2)$, and we have taken all vector fermions to be 
heavier than the top. Note that our definition of $t$ differs 
by a factor of 2 from that used by some authors. For the new fermions, the $\beta$-functions 
are given for the singlet (doublet) type vectors. For simplicity, we have put all the SM Yukawa 
couplings equal to zero except for that of the top quark. This hardly changes our conclusions. 

\begin{figure}
\vspace{-15pt}
\centerline{
\rotatebox{270}{\epsfxsize=5cm\epsfbox{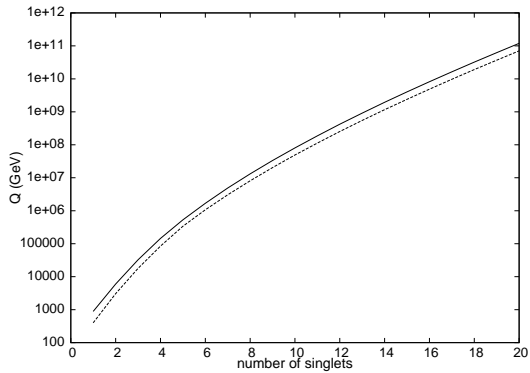}}
}
\caption{The energy scales where (i) the scalar quartic couplings hit the Landau poles (assuming 
one-loop RG equations to be still valid) [upper curve] and (ii) at least one of the scalar couplings 
ceases to be perturbative ($\leq 4\pi$) [lower curve].}
   \label{fig2}
\end{figure}

The main feature of the above equations that all the three scalar quartic couplings, namely, 
$\lambda$, $\tilde\lambda$, and $a$, simultaneously hit their Landau poles; this is because of the 
way their RG equations are coupled. We start with $\tilde\lambda (Q^2=m_Z^2) = 0$, but the result 
is insensitive to the precise boundary condition. The other boundary conditions are fixed by the 
Higgs boson mass and the Veltman conditions. The singlet Yukawa couplings $\zeta_i$ remain 
perturbative for the entire domain. In fact, the DM phenomenology, like the DM-nucleon cross-section, 
does not depend on the precise value of $\zeta_i$, which are fixed only by the singlet Veltman condition.

It might appear counterintuitive that with the increase in $N$, the number of singlets, the 
couplings blow up at a slower rate. However, note that the initial value of $a$ depends on $N$;
while $a^2 (Q^2 = m_Z^2)$ goes down as $1/N^2$, the contribution to $\beta_\lambda$ increases 
only as $N$, so the blow-up is slower. 

Again, one must not read too much in the exact position of the Landau pole for at least two reasons:
first, we do not know as yet how stable the Veltman conditions are if we take higher-order 
quantum corrections, and second, one should not use perturbative $\beta$-functions when at least 
one of the couplings become large. While we have no control over the first issue, except 
dimensional arguments that the higher-order corrections should be small, the second point is
taken care of if we focus not only on the Landau poles but the points where at least one of the 
couplings goes nonperturbative, i.e., above $4\pi$. As expected, this scale is somewhat below 
the Landau pole, as can be seen in Fig.\ \ref{fig2}. Note that the gauge quantum numbers of the 
vector fermions are of no consequence for this plot, as is evident from Eq.~(\ref{all-rge}).

The singlet scalar DM interacts with the SM particles (excluding the vector fermions) only 
through the exchange of the SM Higgs boson $h$ \cite{barger-real}, either by $gg\to h \to SS$ 
or $q\bar{q}\to h \to SS$. Thus, the DM-nucleon scattering cross-section is entirely 
spin-independent. Note that the sensitivity for detection goes down with $m_S$. 

Further tweaks on the model are possible. For example, one can think of mixing between vector
and chiral fermions, consistent with the flavor physics observables. This might open up a
decay channel for the singlet(s), like $S\to f\bar{f}$ (where the stationary state $f$ 
is a mixture of vector and chiral states), and spoil the candidature of the singlet
for DM. This, however, does not invalidate our main conclusion of the necessity of singlet scalar(s) 
and vector fermion(s) for a successful solution of the fine-tuning problem. 
The mixing, of course, can be adjusted to be so tiny as to make the singlet(s) stable on a cosmological 
time scale, but this smells of fine-tuning and goes against the philosophy of this article, 
so we do not recommend this solution. 
One can also do away with the quark doublet altogether, and just think of a lepton doublet $(N,E)$, with
$m_E > m_N$, so that $N$ can either be stable or decay to SM leptons through charged-current 
processes. Even a single heavy neutrino can do the job of solving the fine-tuning problem of the singlet, 
and in fact will be welcome because the decay $S\to\gamma\gamma$ through a heavy fermion loop will 
be automatically absent. 
It is also possible to build
a multicomponent singlet model without the $O(N)$ symmetry, 
but impose further discrete symmetries so that one or more singlets do not couple with the 
vector fermions but the rest do. 

The singlets can even be produced at the LHC, either by gluon-gluon fusion through a vector 
fermion loop, or through radiation off a vector quark. For the former process, one might tag with 
an associated monojet or single photon; for the second, there will be a missing energy signature. 
However, the rates are expected to be quite small for such massive scalars, even at the 
LHC upgrade.

\section{Summary}

We have shown how the fine-tuning problem and the existence of cold dark matter can be linked 
if one assumes an underlying but yet-to-be-discovered symmetry that cancels, or at least 
softens, the quadratic divergence of scalar masses. While the cancellation of quadratic divergences 
for the SM Higgs can be performed by more scalar degrees of freedom, like one or more singlets, 
the same for the singlets necessitates the introduction of vector fermions. Thus, the minimal 
enhancement of the SM that can successfully address both the fine-tuning problem and the existence 
of cold DM is one with one or more singlet scalars that do not mix with the SM doublet, and 
vector fermions, either singlets or doublets under $SU(2)$. 

This model is internally consistent too. The singlet scalars do not mix with the SM doublet because
(i) their self-coupling becomes so large and nonperturbative that the theory hits the Landau pole
at the electroweak scale itself; (ii) for more than one singlets with an $O(N)$ symmetry, there are
$N-1$ Goldstones in the theory, resulting in a large invisible decay width of the SM Higgs. 
The vector fermions do not couple with the SM Higgs, but the stationary states, in particular the 
third generation states, might contain a small admixture of vectorial fermions unless prohibited 
by some discrete symmetry. Such vector fermions and singlet scalars also successfully evade 
the constraints coming from the oblique parameters. 

With these new terms in the potential, the scalar couplings evolve in a different way. With more
positive contributions in the respective $\beta$-functions, all the scalar quartic couplings 
blow up simultaneously somewhere below the Planck scale. The exact position of the Landau pole 
depends on the number of singlet scalars; for one scalar, this is even below 1 TeV, so a 
consistent picture should have at least two real singlets (or one complex singlet). Thus, 
there must be some new physics beyond that point, which may very well be responsible for the 
symmetry. In fact, if the Landau pole is at about 10-100 TeV, the fine-tuning problem is never
that severe, maybe only one or two orders of magnitude higher than what we have for supersymmetric models,
but still it will be aesthetically pleasing to have a way out of the problem.  

There are at least two free parameters of the model apart from those fixed by the Veltman 
conditions: the singlet scalar mass and the vector fermion mass. While the scalar mass can be 
constrained by the relic density of the DM and the DM-nucleon scattering cross section, 
the only constraint on the fermion masses comes 
from non-observability at the LHC. A way to rule this model out might be a nonzero spin-dependent cross-section 
in DM-nucleon scattering. However, we emphasize again that the motivation for this model is to 
supply a minimal extension of the SM to successfully address the hierarchy problem, and the 
appearance of a scalar DM candidate should only be taken as a spin-off.

\acknowledgments
IC was supported by a research fellowship of CSIR, Government of India. 
AK was supported by CSIR, Government of India (project no.\ 03(1135)/9/EMR-II), 
and also by the DRS programme of the UGC, Government of India.

\end{document}